\documentclass{emulateapj}

\slugcomment{{\it Astrophysical Journal} accepted, 2005 July 11}

\shorttitle{{\it GALEX} Ultraviolet Flare}
\shortauthors{ROBINSON, et al.}

\begin{document}


\title{{\it GALEX} Observations of an Energetic Ultraviolet Flare on the dM4e Star GJ~3685A}


\author{
Richard D. Robinson,\altaffilmark{1}
Jonathan M. Wheatley,\altaffilmark{2}
Barry Y. Welsh,\altaffilmark{2}
Karl Forster,\altaffilmark{3}
Patrick Morrissey,\altaffilmark{3}
Mark Seibert,\altaffilmark{3}
R. Michael Rich,\altaffilmark{4}
Samir Salim,\altaffilmark{4}
Tom A. Barlow,\altaffilmark{3}
Luciana Bianchi,\altaffilmark{5}
Yong-Ik Byun,\altaffilmark{6}
Jose Donas,\altaffilmark{7}
Peter G. Friedman,\altaffilmark{3}
Timothy M. Heckman,\altaffilmark{5}
Patrick N. Jelinsky,\altaffilmark{2}
Young-Wook Lee,\altaffilmark{6}
Barry F. Madore,\altaffilmark{8}
Roger F. Malina,\altaffilmark{7}
D. Christopher Martin,\altaffilmark{3}
Bruno Milliard,\altaffilmark{7}
Susan G. Neff,\altaffilmark{9}
David Schiminovich,\altaffilmark{3}
Oswald H. W. Siegmund,\altaffilmark{2}
Todd Small,\altaffilmark{3}
Alex S. Szalay,\altaffilmark{5} and
Ted K. Wyder\altaffilmark{3}}

\altaffiltext{1}{Institute for Astrophysics and Computational Sciences,
Catholic University of America, 200 Hannan Hall, Washington, DC 20064; robinson@opal.gsfc.nasa.gov}

\altaffiltext{2}{Experimental Astrophysics Group, Space Sciences Laboratory,
University of California, 7 Gauss Way, Berkeley, CA 94720;
wheat@ssl.berkeley.edu, bwelsh@ssl.berkeley.edu}

\altaffiltext{3}{California Institute of Technology, MC 405-47, 1200 East
California Boulevard, Pasadena, CA 91125}

\altaffiltext{4}{Department of Physics and Astronomy, University of California,
Los Angeles, CA 90095}

\altaffiltext{5}{Center for Astrophysical Sciences, The Johns Hopkins
University, 3400 N. Charles St., Baltimore, MD 21218}

\altaffiltext{6}{Center for Space Astrophysics, Yonsei University, Seoul
120-749, Korea}

\altaffiltext{7}{Laboratoire d'Astrophysique de Marseille, BP 8, Traverse
du Siphon, 13376 Marseille Cedex 12, France}

\altaffiltext{8}{Observatories of the Carnegie Institution of Washington,
813 Santa Barbara St., Pasadena, CA 91101}

\altaffiltext{9}{Laboratory for Astronomy and Solar Physics, NASA Goddard
Space Flight Center, Greenbelt, MD 20771}



\begin{abstract}

The Galaxy Evolution Explorer ($\it GALEX$) satellite has obtained
high time resolution ultraviolet photometry during a large flare on
the M4 dwarf star GJ~3685A. Simultaneous NUV (1750 - 2800 \AA) and
FUV (1350 - 1750 \AA) time-tagged photometry with time resolution
better than 0.1 s shows that the overall brightness in the FUV band
increased by a factor of 1000 in 200 s. Under the assumption that
the NUV emission is mostly due to a stellar continuum, and that the
FUV flux is shared equally between emission lines and continuum,
then there is evidence for two distinct flare components for this
event. The first flare type is characterized by an exponential
increase in flux with little or no increase in temperature. The
other involves rapid increases in both temperature and flux. While
the decay time for the first flare component may be several hours,
the second flare event decayed over less than 1 minute, suggesting
that there was little or no confinement of the heated plasma.

\end{abstract}




\section{Introduction}
UV Ceti stars exhibit many of the activity phenomena observed on the Sun, such as flares,
dark spots and variable emission from the chromosphere and corona \citep{tho04}.
Such surface activity is linked to
their magnetic field strength and surface coverage, which can be several times larger
than that found on the Sun.
When quiescent, UV Ceti flare stars are low luminosity ($M_V > 8$), low
temperature (T~=~2500~-~4000~K) M-type dwarfs. During outbursts, these stars brighten
dramatically over timescales of a few seconds to hours at all wavelengths from X-ray to
radio \citep{hai91}.

The vast majority of stellar flare observations involve photometric monitoring using the Johnson
UBV filters \citep{lacy}. Normally this involves high speed sampling using a single
transmission filter. Occasionally,
however, observations are obtained in multiple filters by sequencing using time steps of a few
seconds to about 1 minute. This technique has the advantage of obtaining flare colors, allowing estimates
of temperatures and sizes, and is reasonably valid during the long duration decay phase of the events. However,
during the impulsive phase the fluxes can change on time scales of seconds, so the color data becomes
highly uncertain. A second limitation is that the flare energy distribution peaks at wavelengths
shorter than 4000\AA\ \citep{vdo}, so that the Johnson photometry only samples the wing of this distribution,
which is contaminated by the stellar photosphere.

The {\it GALEX} satellite \citep{mar05} provides the opportunity for obtaining vastly improved photometric
measurements of stellar flares. Using a dichroic beam splitter and two photon counting detectors, this
telescope is able to simultaneously monitor a 1.25$^o$ field of view in the FUV (1350 - 1750 \AA) and
NUV (1750 - 2800\AA) regions with 5 arcsec and 6.5 arcsec angular resolution respectively. This allows the accurate determination
of UV fluxes and colors at a time resolution which is limited only by the count rate and the
required S/N values.  In this paper we describe a serendipitous
observation of a large flare seen by the {\it GALEX} telescope
on the dM4e star GJ~3685A. Some of the physical properties of this event are discussed
and we develop
a schematic model which is compared with models developed from other flare observations.

\section{Observations and Data Reduction}

The flare event on GJ~3685A was detected on 2004 April 20 during a
simultaneous FUV and NUV observation of the Medium Imaging Survey
(MIS) field MISDR1\_13062\_0283. The exposure began at 19:42:06 UT
and lasted for 1244s. At present, the standard $\it GALEX$ Data
Analysis Pipeline is only designed to produce a calibrated image of
the field and thus does not retain timing information. To construct
a photometric time sequence, we began with the original time-tagged
photon lists and integrated all photons within a 1.8 arcmin aperture
centered on the source. The background count rate was measured
within an annulus extending from 1.8 to 2.4 arcmin around the star,
during the first 500 s of the observation, in order to avoid
contamination by scattered light when the flare was at its
brightest. The FUV and NUV background rates are $3.2$ cts s$^{-1}$
and 38.5 cts s$^{-1}$, respectively.

The {\it GALEX} microchannel plate detectors exhibit a local
non-linearity of 10\% affecting point sources with input count rates
of 90 cts s$^{-1}$ in FUV and 470 cts s$^{-1}$ in NUV \citep{mor05}.
In this paper, we use Morrissey's published local non-linearity
corrections, which have been calibrated with white dwarf standard
stars up to 3300 cts s$^{-1}$ in FUV and 8400 cts s$^{-1}$ in NUV.
The GJ 3685A flare exceeded this calibrated range in the FUV for a
few seconds during the brightest parts of the flare: we have omitted
these data from the analysis because the uncertainty in the true
count rate is unknown in this high count-rate regime.

\begin{figure}[htbp]
\includegraphics[angle=90,scale=.8]{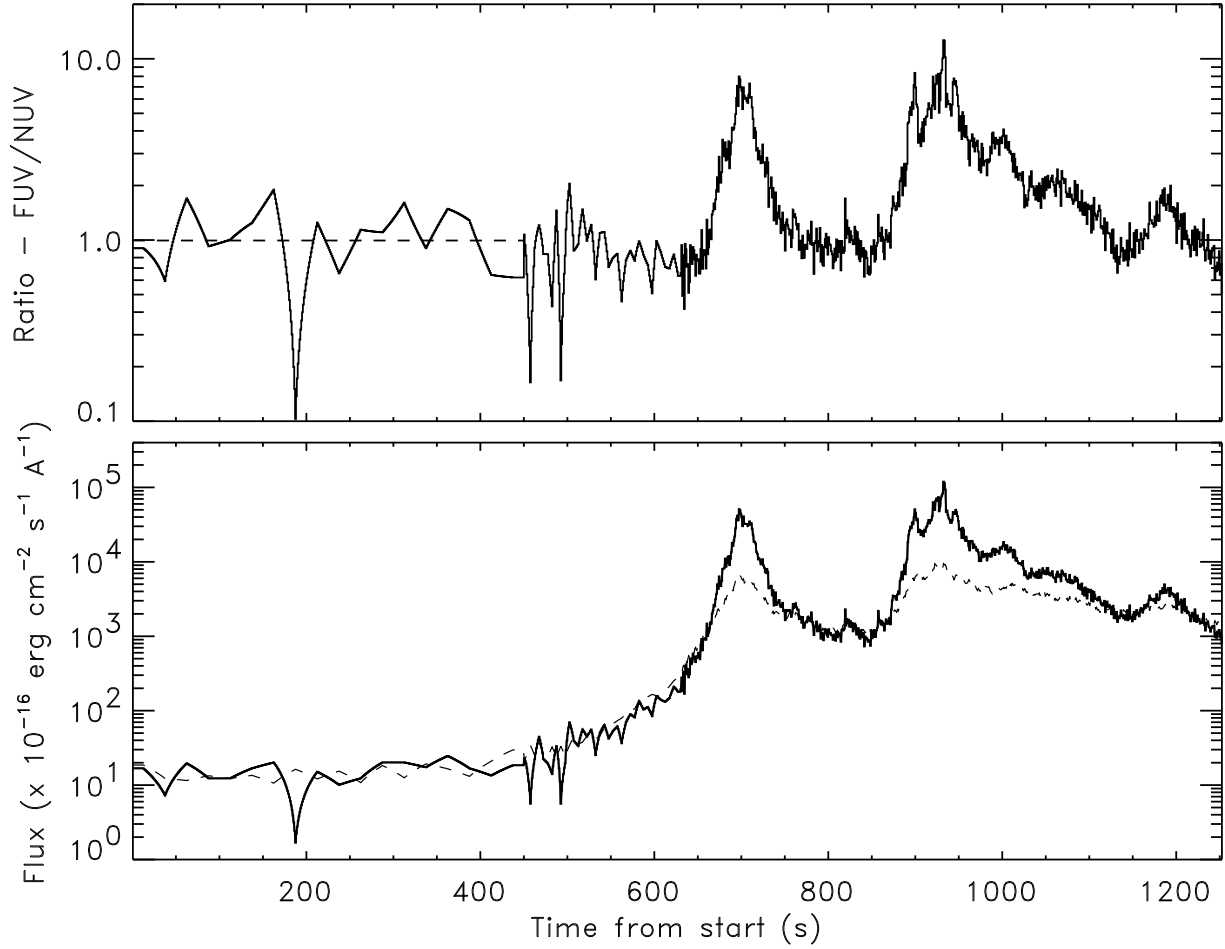}
\caption{(Top) Ratio of FUV to NUV fluxes. In this and all other
figures, data are omitted during the time intervals in which the FUV
count rate exceeds the maximum calibrated count rate of 3300 cts
s$^{-1}$. (Bottom) Calibrated fluxes from the {\it GALEX} FUV
(solid) and NUV (dashed) channels for the entire observation of 2004
April 20. Data have been binned into 25 s intervals for 0-450 s, 5 s
intervals between 450 and 630 s and 1 s intervals for times greater
than 630 s.} \label{ratio}
\end{figure}

The results of the analysis are given in Figure \ref{ratio},
which shows both the flux variations and the
ratio of the FUV to NUV fluxes. In these plots the photon count rates have been
converted to fluxes (in erg cm$^{-2}$  s$^{-1}$ \AA$^{-1}$) using the conversion factors
of $1.4\times10^{-15}$ and $2.6\times10^{-16}$ for the FUV and NUV, respectively \citep{mor05}. Initially, the star showed
a `quiescent' flux of $\sim 10^{-15}$ erg cm$^{-2}$ s$^{-1}$ \AA$^{-1}$ and a
flux ratio (or color) of 1.
The start of the stellar flare is lost in the noise of the background. A reasonable
estimate seems to put the start somewhere
between 450 and 500 s and it is definitely underway in both the NUV and FUV
bands by 500 s. Between 500 s and 630 s the flux in both bands increases
exponentially, with an e-folding time of $\sim$60 s. During this time the
average color remains constant, at a value of near 0.8.

At 650 s the rate of flux increase jumps dramatically, with
e-folding times of 23 s in the NUV and 12 s in the FUV, and the
FUV/NUV ratio increases from $ < 1$ to more than 6.  Both
temperature and FUV/NUV ratio reach a peak at 700 s. During this
peak period, we have omitted the data because the FUV non-linearity
correction is uncertain.

After the peak, which only lasts for about 20 s, the flux and
FUV/NUV ratio rapidly decrease. Note that the time variations of
both the color and the flux for the 100 s centered at the flare peak
are symmetrical about the peak, i.e. the rise times and decay times
are about the same. By 750 s the color ratio is again near 1 and the
flux settles into a slow decline which is interrupted at 865 s by a
second flux enhancement. This second enhancement is somewhat more
complicated than the first, but follows the same basic trend of
rapid flux increase accompanied by a rapid increase in FUV/NUV, this
time to values in excess of 10. The decay phase for the second
enhancement is slower and more complex than was the case for the
first enhancement. By the end of the observation the flux and color
had returned to the levels seen prior to the second enhancement,
though the flux is still well above the `quiescent' level.

\section{Analysis}

\subsection{Pre-flare Activity}

On 2004 March 2 {\it GALEX } obtained a 110 sec `quiescent'
observation of GJ~3685A which showed a flux of
$1.7\pm0.5\times10^{-16}$ erg cm$^{-2}$ s$^{-1}$ \AA$^{-1}$ in the
FUV and $3.0\pm0.3\times10^{-16}$ erg cm$^{-2}$ s$^{-1}$ \AA$^{-1}$
in the NUV, giving a color ratio of $0.57\pm0.19$. On 20 April,
before the onset of the flare event the average `quiescent' flux was
$1.54\times10^{-15}$ erg cm$^{-2}$ s$^{-1}$ \AA$^{-1}$ in the NUV
and $1.52\times10^{-15}$ erg cm$^{-2}$ s$^{-1}$ \AA$^{-1}$ in the
FUV band, resulting in a flux ratio of 1. Since the FUV/NUV flux
ratio is directly related to the effective emission temperature (see
section \ref{model}) we see that preceding the flare the stellar
emission had not only increased in flux by a factor of 5 in the NUV,
but also increased in temperature.

One possible explanation is that the pre-flare enhancement results
from an increased level of microflaring activity. To test this
notion, we performed an analysis on the first 400~s of the data set
using the binning technique described by \citet{rob95,rob99}.
Individual microflares can be detected by binning the time sequence
and searching for times when the counts per bin are significantly
higher than those expected from random chance. The maximum
visibility occurs when the binning factor is near the lifetime of
the event.

A distribution of small events which are not individually time
resolvable can be determined by comparing the distribution of count
rates with a reference distribution from a non-variable source, as
described in Robinson (1999). This method reveals that when the
binning factor is much smaller than the flare lifetime, then event
noise will dominate and both distributions will be Poisson. However,
as the binning factor approaches the flare lifetime, the microflares
show up as an enhancement of the high count rate tail of the
distribution.

Examining the time sequences for binning factors ranging from 1 s to
20 s for the GJ 3685A flare data shows no strong evidence for either
individual or unresolved flare events, while a regression analysis
showed that the level of activity remained statistically constant
prior to flare onset. A problem, however, is that the background
levels are large compared to the stellar signal. For example, in the
NUV the estimated background is 38.5 counts s$^{-1}$, while the
stellar signal was only 7.5 counts s$^{-1}$. Thus, noise from the
background would swamp all but the largest microflare events.

\subsection{Empirical Modeling}
\label{model}

Since we have no spectroscopic information about this flare it is
not possible to perform any type of detailed modeling or radiative
transfer calculations on these UV data. However, some good
qualitative and semi-quantitative insights can be obtained by
assuming black-body emission. The reader should be aware that this
assumed model is only one of several possible flare model scenarios.
However, this simple theoretical approach has been widely used in
the interpretation of optical photometry of flare events and yielded
some important insights into flare phenomena
\citep{kahler,vdo,hawley03}. We have a definite advantage in these
UV observations since the photospheric emission from the star is so
small that it can be neglected.

\begin{figure}[htbp]
\includegraphics[angle=90,scale=.8]{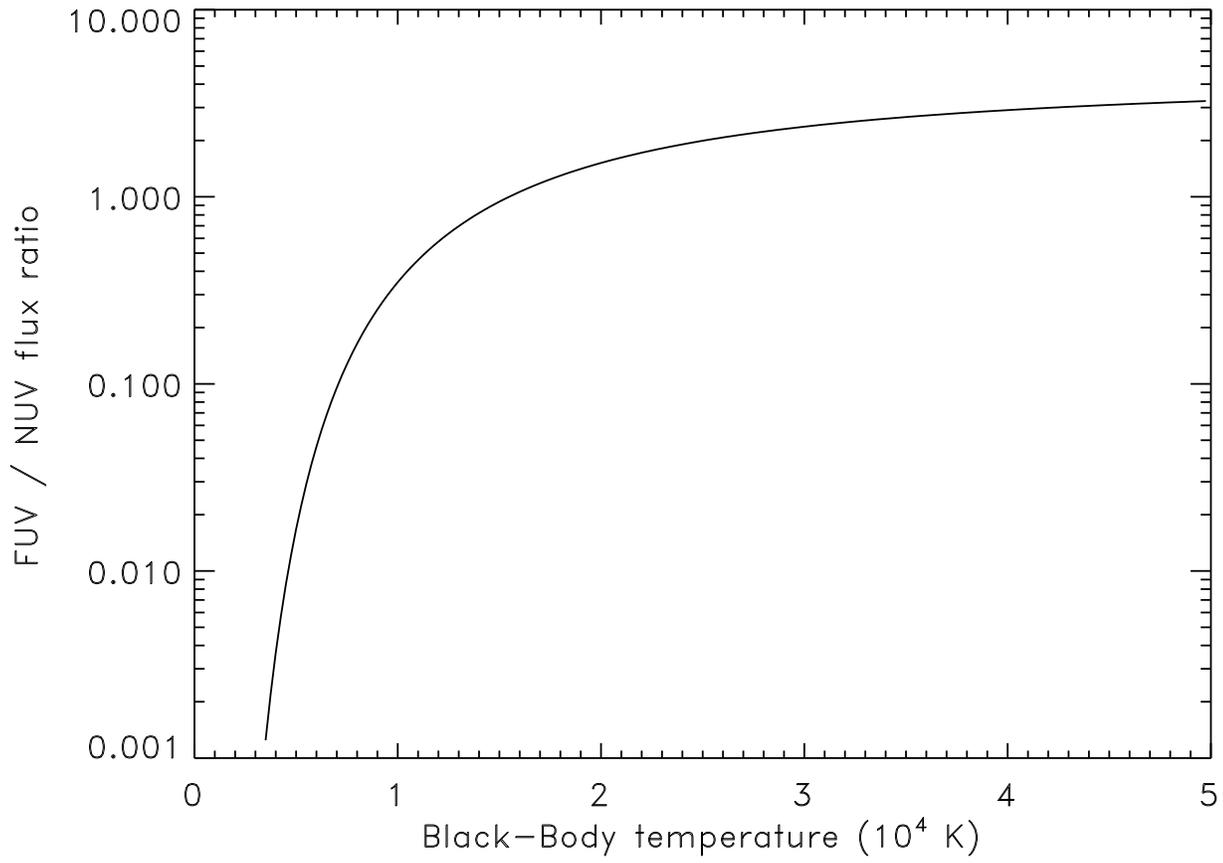}
\caption{Relation between black body temperature and the FUV/NUV flux ratio
determined by convolving the appropriate black body distribution with the
{\it GALEX} FUV and NUV filter effective area curves. }
\label{tempcal}
\end{figure}

The first step in our simplified analysis is to determine the
relation between the black body temperature and the measured FUV/NUV
flux ratio. We have calculated the black body spectrum for a number
of different temperatures, multiplied by the effective area curves
for the {\it GALEX} FUV and NUV filters and then divided by the
total effective area for each filter. This gives the expected flux
for a black body, which can be directly related to the average flux
deduced from the measured count rates. The relation between
temperature and flux ratio is shown in Figure \ref{tempcal}. Note
that the flux ratio increases dramatically between 4000 and 19000 K
(when the peak of the black body distribution falls within the FUV
bandpass) and then increases much more slowly toward high
temperatures. The ratio eventually reaches a value of just over 3 at
temperatures of 50,000 K. This immediately shows that our black-body
assumption is not totally valid, since we measured ratios during the
flare peaks in excess of 6. This is not a surprise, since the FUV
wavelength region has numerous strong emission lines (e.g. C II, C
IV, Si II, Si IV, etc)  which can dominate the continuum, especially
during `quiescent' periods outside of major flare events
\citep{vdo}.

Assessing the relative contributions from continuum and/or emission
lines at near UV wavelengths is still an open problem for flare
research \citep{haw92}. We note that the GALEX NUV channel is not
significantly contaminated by the (expected) strong MgII emission
lines at 2800\AA\ due to the low transmission of the GALEX NUV
instrument at this wavelength. However, previous flares have shown
an (albeit far less) increase in the NUV emission from the FeII line
multiplet at 2600\AA\ \citep{hai87}, whereas \citet{but81} have
observed a significant rise in the near UV continuum during a flare
on the star Gl 867A. For the present case of the flare on GJ 3685A,
it is difficult to imagine how line emission could be the sole major
contributor to the more than 5 UV magnitude increase in NUV flux
observed during the flare. Thus, in determining the effective
temperature from the FUV/NUV flux ratio, we {\it arbitrarily} assume
that approximately half of the FUV flux was from emission lines and
the other half from black-body emission. This is consistent with the
time-resolved spectroscopy of \citet{hawley03}, where the FUV
line-to-continuum flux ratio is 0.8, during the brightest flare on
AD Leo reported in that paper. Furthermore, \citet{haw92} also
provide support for the notion that the {\it GALEX} NUV band is
dominated by the star's continuum.

\begin{figure}[htbp]
\includegraphics[angle=90,scale=.8]{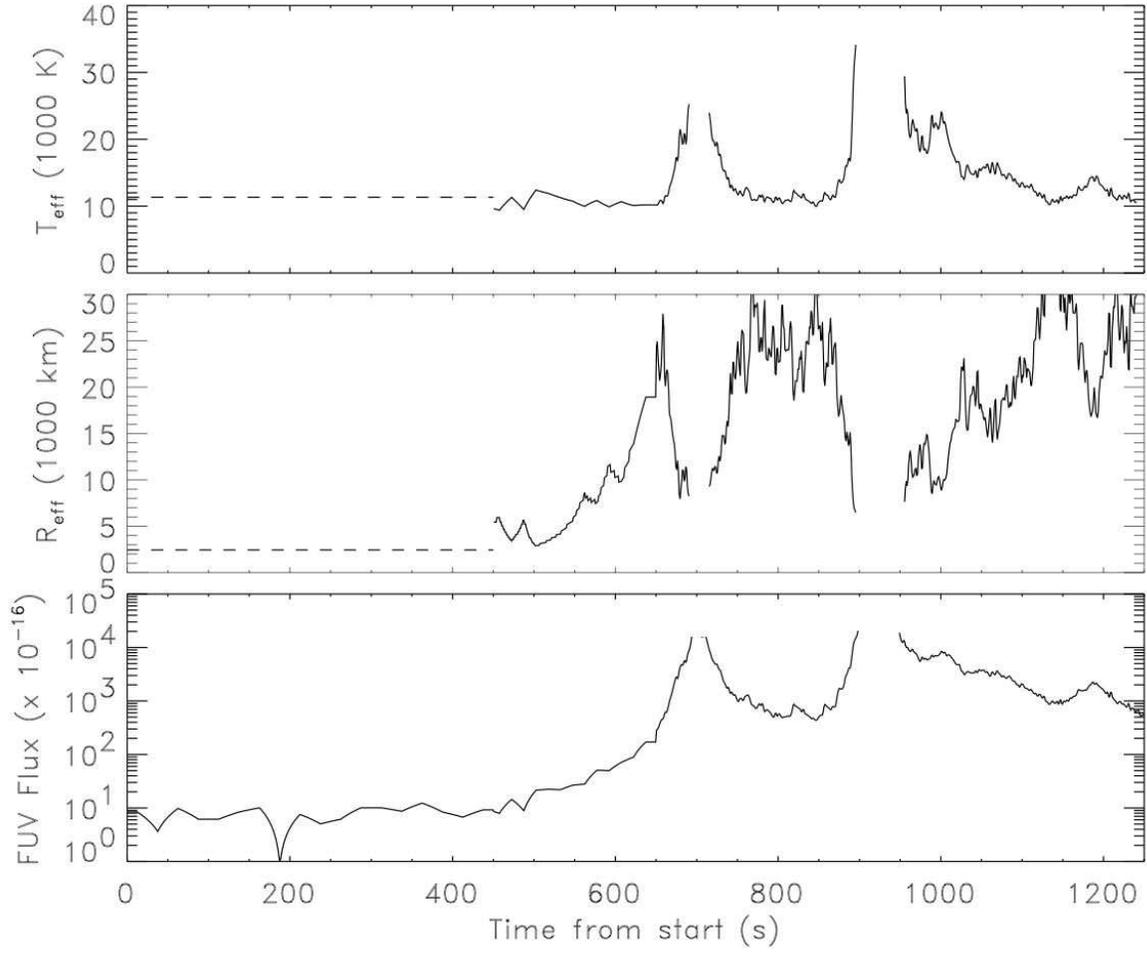}
\caption{(a) Effective black body temperatures derived from the
measured FUV/NUV flux ratio using the calibration presented in
Figure \ref{tempcal}. Preflare activity has been estimated using an
integration over the first 400 s. Data between 400 s and 630 s is
deduced using 5 s bins, while later data uses 1 s bins. (b)
Effective radius of the source deduced from the observed flux, as
described in the text. Data for the first 630 s has been binned into
5 s intervals, later data has been binned into 1 s increments. (c)
Measured FUV flux, binning as in (b).} \label{tempvar}
\end{figure}

A black body of given temperature has a well defined emission. Thus,
having determined a temperature we can define the effective stellar
surface coverage required to produce the observed NUV flux at the
earth, assuming a distance to the star of 14.6 pc \citep{ast98}. The
results of the calculations are presented in Figure \ref{tempvar},
where we have converted the flare surface area to an effective
source radius. The calculated temperature in the pre-flare phase was
about 12,000 K with a deduced scale of about 3500 km. The exact
values are uncertain because this phase is dominated by the emission
lines in the FUV and the actual structure is most likely composed of
small elements spread over a large area. When the flare starts the
effective temperature drops to about 10,000 K and the radius
increases linearly from 3,500 km at 500 s to about 25,000 km at 650
s, suggesting an initial expansion velocity of at least 140 km
s$^{-1}$. The deduced temperature increases dramatically at 660 s,
becomes undefined near the flare peak and then rapidly decreases
back to a value of about 10,000 K. During this time the deduced
radius decreases to 7,000 km and then increases back to
25,000-28,000 km, about the same value as that seen before the large
temperature increase. The temperature and area then remain
relatively constant until the start of the second temperature
enhancement at 880 s, where the area again dramatically decreases
and then slowly recovers to a value of about 32,000 km.

\begin{figure}[htbp]
\includegraphics[angle=0,scale=.8]{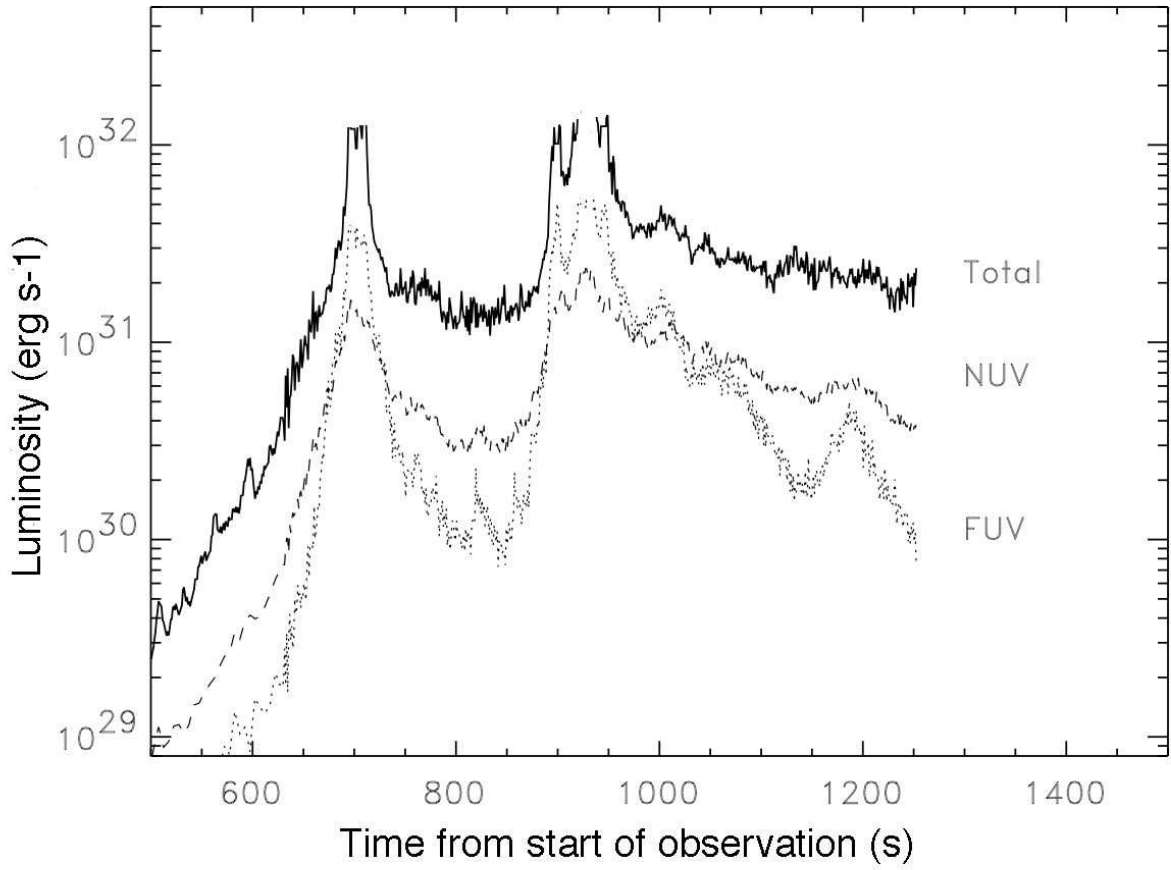}
\caption{Estimated luminosity, assuming black body emission at the
temperature and source size presented in Figure \ref{tempvar}.}
\label{energy}
\end{figure}

Having estimated a temperature and area for the flare it is possible
to determine the total optical and UV energy by integrating under
the appropriate black body curve. The result of this calculation is
shown in the Figure \ref{energy}. The overall appearance of this
event is two short duration bursts at 670-730 s and 880-970 s
superimposed on a more slowly varying event which started at 450 s
and which was slowly decaying at the end of the observation.  The
flare is obviously very energetic, with luminosities reaching more
than $10^{32}$ erg s$^{-1}$ and a total integrated energy in excess
of $10^{34}$ erg. This is comparable to the largest recorded M-dwarf
flare outbursts, such as the 1985 flare on the star AD Leonis
\citep{hawley91}. Another way to put this into perspective is to
note that the total (optical plus UV) luminosity emitted from the
photosphere of a M4 star is on the order of $2.0\times10^{31}$ erg
s$^{-1}$. Thus, at the peak the flare would outshine the star by
more than a factor of 10.

\section{Discussion }

Assuming that our black body emission model discussed in Section 3.2
provides a reasonable description of the flare event on GJ 3685A
(and given the caveat that there may be other models that can
provide an alternate interpretation of the data that we have chosen
not to explore due to the lack of spectroscopic information on this
flare), we can now postulate a schematic model for the event which
satisfies all of the observations and is consistent with
observations of flares seen on other stars. The flare itself appears
to be composed of two distinct types of events, which we will refer
to as type A and type B. The flare begins with a type A event, which
is characterized by a FUV/NUV flux ratio that is nearly the same as
that seen in the pre-flare activity (plage) and remains constant as
the flux increases. We conclude that the spectrum  has a very
plage-like appearance and that the source structure, and possibly
the heating processes, are also very plage-like. We postulate that
this phase was initiated by a magnetic reconnection somewhere within
a highly stressed active region. This initial energy release
triggers reconnection events in adjacent magnetic structures,
leading to an avalanche \citep{lu,carb} which propagates throughout
the region at a velocity of at least 140 km s$^{-1}$, which probably
represents the local Alfven velocity. The increase in flux is then
simply the result of expanding area coverage.

At about 670 s (roughly 3 minutes after the start of the flare) the disturbance
which is responsible for flare A intersects a highly unstable magnetic structure
and triggers an explosive release of energy, resulting in the first type B flare
(B1), which occurs between 670 s and about 730 s. The rapid increase
in temperature and flux is compatible with a chromospheric evaporation
model \citep{hai91}, in which a large amount of energy (probably in the form of energetic particles)
is suddenly released near the top of a magnetic loop complex and propagates down
towards the photosphere. When it strikes the denser atmospheric layers it
impulsively heats the material, which then expands back into the loop. Normally, this type of
event will have a long lasting decay as the heated material cools through conduction and radiation.
In this case, however, the short duration of the decay suggests that the energy release
was sufficient to cause a disruption of the confining loops \citep{reale}. The source size for flare B1
is also significantly smaller than that of flare A, since the deduced flare area
decreases rapidly when the emission from B1 dominates that from flare A.

The structure responsible for flare B1 is probably near the edge of the
active region associated with flare A, since the expansion of flare A
ends during flare B1. Apparently, all of the available structures which can
support flare A emission have been activated by that time.
It is possible that the
disturbance responsible for flare A could continue into regions outside of the
initial active region. More likely, the energy release initiating flare B1
also generated an Alfvenic shock, similar to the Morton waves seen on the Sun
\citep{athay}.
This disturbance propagates to an unstable magnetic structure in a nearby
active region and triggers a second B type flare (flare B2) which starts
at around 880 s. This flare is both stronger and more complex than B1, indicating
more complex field configuration. It also shows a better developed
thermal decay, implying that at least some of the hot plasma is confined
in the loops. Note also that this is an isolated event, without an accompanying
type A event, since there was no increase in source size prior to the onset of
the temperature enhancement.

By the end of the observation it appears
that flare B2 has completely faded and flare A is again visible. Unfortunately,
we do not have much information about the decay phase of flare A. However, from
the fact that the integrated flux at the end of the observation at 1250 s is very
similar to that at 800 s, after the decay of flare B1, suggests that it may take
an hour or more for flare A to decay to pre-flare levels.

\section{Conclusions}

We report on a large flare from the dM4e star GJ~3685A which was
simultaneously observed in the NUV and FUV  by the {\it GALEX}
satellite. Under the assumption of a blackbody emission model in
which half of the FUV flux arises from line emission and half from
continuum, and that the NUV flux arises solely from an increase in
the continuum, there is strong evidence for two distinct classes of
flares during this event. The first (type A) is characterized by an
exponential increase in flux with little of no change in emission
temperature, as measured from the FUV/NUV flux ratio. This behavior
is compatible with an avalanche model. The fact that the flux ratio
during this phase was very similar to that seen in the pre-flare
plage emission also supports the idea that plages are heated by
microflares, as first proposed by \citep{parker}.

The second class (type B) is characterized by an impulsive increase
in both flux and temperature. This class is consistent with a
classical explosive event and may be related to the solar two ribbon
flare. Two flares of this type were seen during the event. The first
(flare B1) probably originated in the same active region as the type
A flare. The lack of a well developed decay phase suggests that the
confining magnetic structure was disrupted. The second flare (B2)
may have originated outside of that region, being triggered by a
disturbance formed during the flare B1 energy release.

The 2004 April 20 flare began when GJ~3685A was already in a state
of enhanced activity in which both the flux and temperature are
substantially higher than in a previous {\it GALEX} observation on
2004 March 2. It is unclear whether this enhancement is the result
of coronal variations, or arises from the rotation of an intense
active region onto the visible surface of the star, or if it comes
from some pre-flare energy release. In order to obtain better models
of future dMe star flare events we recommend using the
low-resolution spectroscopic grism mode of {\it GALEX}, instead of
the imaging photometric mode. Such future observations would provide
better insights into the relative contributions from the UV
continuum and/or line emission during these large releases of
stellar energy.

\acknowledgments {\it GALEX} is a NASA Small Explorer, launched in
2003 April. We gratefully acknowledge NASA's support for
construction, operation, and science analysis for the {\it GALEX}
mission. Financial support for this research was provided by NASA
grant NAS5-98034.










\newpage

\newpage

\newpage

\end{document}